# A VUV light source for enhanced production of metastable krypton and xenon beams

P. R. STOLLENWERK, K.G. BAILEY, D. KOCH[†], P. MUELLER, T.P. O'CONNOR, J. C. ZAPPALA, AND M. BISHOF[*]

*Physics Division, Argonne National Laboratory, Lemont, Illinois 60439, USA*
[†]*Current Address: University of Tennessee Knoxville, Knoxville, Tennessee 37996, USA*
[*]bishof@anl.gov

**Abstract:** We demonstrate excitation of metastable krypton and xenon beams using a vacuum ultraviolet lamp and directly compare the performance of this method to metastable excitation based on a radiofrequency-driven plasma discharge. In our apparatus, lamp-based metastable excitation outperforms the plasma discharge across a wide range of beam flux values relevant for Atom Trap Trace Analysis (ATTA). Moreover, we do not observe significant degradation in lamp performance after over 160 hours of operation. We find that lamp-based excitation is particularly advantageous at the smallest and largest beam fluxes tested, demonstrating the utility of this approach both for improving krypton ATTA and for enabling the detection of radioactive xenon isotopes using ATTA. Finally, we demonstrate an additional enhancement to lamp-based metastable excitation efficiency and stability by applying an external magnetic field.

## Introduction

The generation of vacuum ultraviolet (VUV) light (photon energies of 6 to 12 eV) is required for a broad spectrum of applications such as semiconductor fabrication [1], materials science [2], photochemistry [3, 4], and spectroscopy [5, 6]. One particularly challenging application has been the production of metastable noble gas atoms, where bright, narrow-band VUV light precisely tuned to an atomic transition is needed [7, 8, 9, 10, 11]. Noble gas atoms populated in long-lived metastable state are, themselves, used in a myriad of applications. In particular, laser-cooled and trapped metastable noble gas neutral atoms have been used in fundamental and applied sciences to study Bose-Einstein condensation [12], cold chemistry [13], precision spectroscopy [14], fundamental physics [15], and radiometric dating of ice [16, 17] and water [18, 19] via noble gas radioisotope tracers.

The density, efficiency, or absolute rate of metastable excitations often limits applications that require metastable noble gas atoms. An intrinsic challenge to improving metastable excitation is the large energy difference from the ground to the first excited state (8 to 20 eV). This presents a challenge for direct laser excitation, which was only recently demonstrated in Kr [7, 20]. These pioneering studies are, however, limited to low duty-cycle, pulsed excitation within a small volume of a vapor cell with overall small excitation efficiency. For this reason, applications using metastable noble gas atoms largely rely on non-resonant electron excitation in either a DC or RF-driven plasma discharge [21, 22, 23, 24, 25]. Yet, competing processes within a plasma discharge, such as de-excitation and ionization, depend strongly on gas density and composition and typically limit metastable production in a noble gas beam to an efficiency of $1\times10^{-4}$ or less.

Narrow-band VUV lamps have the potential to overcome these limitations and achieve greater metastable excitation density, efficiency, and absolute excitation rate via resonant excitation to an excited state followed by infrared laser excitation to a higher lying state that preferentially decays to the metastable state (see Figure 1). Previously demonstrated lamp-based sources of VUV light lacked the combination of sufficient power, stability, and operational lifetime required for routine applications. To date, excitation efficiencies achieved with such lamps have remained smaller than those of discharge sources and VUV light transmission through MgF$_2$ lamp output windows has significantly degraded after only a few hours of operation [8, 9, 10].

Here, we present a novel lamp design that overcomes these difficulties and we demonstrate lamp-based metastable excitation in effusive atomic beams of krypton and xenon with overall excitation efficiency and total metastable flux exceeding the performance of an RF-driven discharge. Our lamp is particularly useful in the detection of rare noble gas isotopes using the Atom Trap Trace Analysis (ATTA) technique, where it is essential for metastable atomic beam sources to achieve both high total flux and high metastable excitation efficiency [8, 26].

The benefits of resonant optical metastable production in an ATTA system are significant. As a radio-isotope analysis tool, ATTA stands alone in its complete insensitivity to radioactive background or contamination from other species. The ATTA technique has enabled radio-dating of ice and water samples with isotopic abundances that would otherwise be below the detection threshold of more conventional techniques such as mass-spectroscopy or low-level decay counting [27]. An improved source for metastable krypton (Kr*) on an ATTA instrument would enable a significant expansion in the precision of groundwater analysis critical to wide-spread use of the technique in the most drawn aquifers in the world [28, 29], as well enabling rare ice core studies [8], particularly by virtue of removing cross-sample contamination issues stemming from the plasma discharge approach [30]. Furthermore, the lack of an efficient, high-flux source of metastable xenon (Xe*) currently precludes similar application of ATTA for applications related to monitoring of Comprehensive Nuclear-Test-Ban Treaty (CTBT) compliance, where characterization of the precise ratio of the four relevant radioxenon isotopes ($^{131m}$Xe, $^{133m}$Xe, $^{133}$Xe, and $^{135}$Xe) is crucial. Next generation radioxenon monitoring systems in the International Monitoring System (IMS) are expected to achieve minimum detectable concentrations of ~0.2 mBq/m$^3$ [31]. Equivalently, such atmospheric concentrations correspond to isotopic abundances between 4×10$^{-18}$ and 1×10$^{-16}$. In a xenon ATTA system, the maximum measurement time will be limited by the 9 h half-life of $^{135}$Xe. For comparison, the detection limit of the state-of-the-art krypton ATTA for a 2-4 h measurement is at the 3×10$^{-15}$ level and is limited by background from long-lived cross-contamination isotopes [32]. Therefore, an optical source of Xe*, which avoids cross-contamination, with throughputs 10 to 100 times greater than the current krypton ATTA state-of-the-art, could enable a xenon ATTA system with sensitivities beyond current analysis capabilities in the IMS.

In this work we demonstrate and characterize Kr* and Xe* excitation using a novel VUV lamp design and compare it to a conventional RF-driven discharge source used in state-of-the-art ATTA instruments. Our lamp design leverages external magnetic fields to enhance both output intensity and stability, and, in contrast to previous lamp designs, we find that metastable excitation efficiency does not appreciably degrade after more than 160 hours of lamp operation. We achieve this unprecedented combination of strong VUV output and long lifetime by maintaining a continuous flow of lamp gas away from the MgF$_2$ window surface and isolating the lamp plasma from all surfaces except MgF$_2$ and quartz. For both Kr and Xe we find that the all-optical approach matches or outperforms the RF-discharge source at all beam fluxes and particularly so at higher fluxes where a Xe ATTA system is likely to operate. Notably, we find Xe* production efficiency to be more than an order of magnitude greater than that of Kr* indicating the feasibility of an all-optical source a practical Xe ATTA instrument. Finally, we report on evidence for multifaceted enhancement of lamp operation via application of external magnetic fields.

## Experimental Methods

Metastable production and detection pathways for the two approaches compared in this work, i.e., RF discharge and all-optical excitation, can be described for both krypton and xenon in a single, simplified 5-level energy diagram shown in Figure 1. Each method is applied with the intention of populating the metastable "$|m\rangle$" state, i.e. the N+1 s[3/2]2 state of Kr or Xe (N=4 or 5, respectively). In the all-optical approach, the metastable state is populated via a 3-step process. The first step is accomplished by exciting $|g\rangle \rightarrow |i\rangle$ with VUV photons (123.6 nm for Kr, 147.0 nm for Xe) supplied by a lamp containing either Kr or Xe gas. Simultaneously, a 1.2 W Ti:Sapphire laser pumps the $|i\rangle \rightarrow |e\rangle$ transition using NIR light (819.2 nm Kr, 895.5 nm Xe). Finally, successful transfer to the metastable state $|m\rangle$ occurs according to the branching ratio via spontaneous decay from $|e\rangle$. The discharge approach is described in [21]. Briefly, it relies on electron collisions to populate the metastable state. In the discharge source mode, the RF generator driving the plasma typically operated using 15-20 W of power at around 80 MHz.

In the experiments relying on optical excitation from the microwave lamp, the atomic beam is created by flowing either Xe or Kr into the chamber through a 1.3 cm diameter capillary array of densely packed 2.5 mm length tubes with 25 um diameters and an open area ratio of around 50%. Excitation occurs immediately after the exit of the capillary array with the lamp window positioned perpendicularly to the atomic and IR laser beams (see Figure 2a). The IR beam is counterpropagating to the atomic beam and is slightly focused such that its diameter covers the entire capillary array. In contrast, in the discharge experiments, Xe or Kr is flowed through a 17.5 cm long quartz tube with 1.1 cm inner diameter as the discharge is applied. As excitation occurs before the tube exit, the beam flow cannot be directed with a capillary array without severely degrading metastable production efficiency.

The metastable detection setup is identical for both discharge and optical experiments. Metastable production in the atomic beam is measured by probing the cycling transition in Kr (811.51 nm) or Xe (882.18 nm) with a CW diode-laser tuned to the $|m\rangle \rightarrow |c\rangle$ transition and the resulting laser induced florescence is imaged onto an amplified Si photodiode equipped with an appropriate bandpass filter (55%, 58% transmission for Kr and Xe respectively). A schematic of the setup is shown in Figure 2a. The detection laser has an 8 mm beam diameter and is aligned perpendicularly to the atomic beam 33 cm and 15 cm downstream from the atomic source for the discharge and lamp setups, respectively.

To infer the metastable flux conversion efficiency within the atomic beam we operate in a regime such that the probe laser is tuned to the most abundant isotopes $^{84}$Kr (57% abundance) and $^{132}$Xe (27% abundance) and is more than an order of magnitude above the saturation intensity. Consequently, we derive an effective cylindrical volume of metastable atoms with a maximum uniform scattering rate with diameter of 12 mm after accounting for the effect of the gaussian spatial profile on the saturation intensity. The high saturation parameter allows the average scattering rate per atom to be insensitive to power fluctuations and, convolved with the gaussian intensity profile, it provides sharply defined boundaries of the probed volume. The overlap of the probed volume with the detected volume is estimated analytically to determine an effective solid angle and volume dimensions [33]. Together, the photodetector response, the effective solid angle, and the scattering rate provide an accurate estimation of the total number of metastable atoms present in the calculated overlap volume. Determining the conversion efficiency is therefore reduced to dividing by the isotope abundance adjusted number of atoms in the overlap volume. The total number of atoms from the atomic beam in the overlap volume is calculated following [34] for a single effusive tube source and extended to the capillary array source by integrating the contribution over all micro-tubules. Here we assume an empirically determined temperature of 375 K to estimate the velocity of the beam.

A schematic of the optical excitation lamp and its mounting design is shown in Figure 2b. A discharge is generated in the lamp using an external tesla coil and sustained by a McCarroll microwave cavity [35] operated at 2.45 GHz with up to 200 W of power coupled. The VUV light exits out a 25 mm diameter $MgF_2$ window (5 mm thick) facing perpendicularly to the atomic beam at the exit of the capillary. The lamp is housed inside a Faraday cage and mounted in a reentrant flange enabling the VUV output flux to be adjacent to the atomic beam. The lamp is designed using an outer (19 mm diameter) and slightly recessed (~1 mm) inner quartz (12.5 mm diameter) tube. At the inlet, a glass stem is attached to the outer tube to supply lamp gas and is pumped out through the back end of the inner quartz tube such that the lamp can operate in the continuous flow regime to limit contaminant build up on the window. The flow rate is varied by an electronically controlled stepper valve connected to the input of a turbomolecular pump. The lamp is sealed off from the atomic beam vacuum chamber by pressing the window into a 1.5 mm diameter indium wire on the lamp mount with a threaded cap.

Short operating lifetimes have been a challenge for VUV lamps [8, 9, 10]. It is hypothesized that contact between the plasma and any metal surface creates contaminants that can be subsequently deposited onto the $MgF_2$ window surfaces and quickly reduces its transparency [8, 36]. To circumvent this issue, we rely on the external air pressure to press the polished end of the outer quartz tube against the window, isolating the Viton O-ring and the metallic vacuum chamber from the area of the lamp window exposed to the plasma.

During tests of the xenon lamp metastable production, we discovered that a magnetic field applied to the plasma could both increase the lamp brightness and metastable production as well as coerce the plasma into a more stable mode of operation. In the optimal configuration, a permanent, axially magnetized, cylindrical magnet was placed approximately 6 cm from the center of the lamp window (see Figure 2b) such that the measured field strength is between 20 and 40 G at a distance corresponding to inside the lamp.

## Results and Discussion

### Lamp Characterization

Figure 3 shows the immediate effect of the external magnetic field on the metastable output of xenon. For this comparison, the lamp operates near the empirically determined optimal xenon partial pressure of 6 mTorr with ~140 W of coupled microwave power. This pressure, as in previous work [9, 8], optimizes the self-absorption broadening caused by ground-state noble gases near the exit window against the reduced number of emitters at lower pressures. Without an applied magnetic field, plasma mode stability is a persistent challenge in the low-pressure, high-power regime that is ideal for VUV output. A small addition of a krypton buffer gas (~5 mTorr partial pressure) helps to avoid spontaneous extinguishing of the plasma, however, the addition of more krypton leads to preferential operation in suboptimal plasma modes where maximum brightness does not occur near the exit window. Therefore, the maximum signal achieved without the magnet was hard to maintain as indicated by the orange triangles in Figure 3, making it impractical for applications in ATTA. With the magnetic field present, we visually observe the plasma mode increase in brightness and proximity to the face of the exit window. This correlates with an enhanced signal (blue squares in Figure 3). Furthermore, the magnetic field increases the stability of the plasma mode, allowing it to operate with optimal metastable production for hours without manual adjustment, though the magnitude of the signal is sensitive to the precise position and orientation of the magnet. We observe that the metastable flux is commonly stable to ~5% over a few hours and fluctuates with small drifts in lamp pressure and temperature. We believe that stability could be further improved with active feedback on those parameters.

Because maximum signal at lower pressures requires up to 140 W of microwave power, heating of the lamp is significant, and it can reach temperatures up to 450 K if there is no active cooling to the lamp body. For lamp operation, we cool the cavity by flowing air directly into the cavity and separately onto the base of the lamp at the connection to the vacuum chamber. Without sufficient cooling, heating from the microwave source occasionally causes air leaks to develop in the lamp as the quick connect fittings loosen with temperature changes. We believe that a more aggressive and uniform cooling of the lamp could eliminate these issues and improve the stability of VUV output.

In Figure 4 we plot the average Xe* operating efficiency of the lamp as a function of lamp operation time (i.e., time with an active plasma discharge). Averages are calculated using data collected during stable operation in 22-hour lamp operation time bins. Here, stable operation is defined to be 5 consecutive minutes or more of signal with less than 5% variation from point to point. This definition is used to exclude from the data periods where manual lamp tuning or laser instability occurred. Parameters during stable operation were optimized for maximum signal and the chamber pressure was kept at a constant $1\times10^{-5}$ Torr to avoid flux and pressure dependent changes in efficiency. Typically, the lamp operated with 100-140 W of coupled microwave power and a partial pressure between 6 and 10 mTorr. Discovery of the magnetic enhancement was not made until after 22 hours of lamp operation time, therefore the first time-interval data point in Figure 4 is distinguished to reflect this fact.

No significant degradation of the signal is observed over the 166 hours of operation. The maximum signal recorded over each 22-hour interval is also plotted in Figure 4 and corroborates the lack of any significant trend. Nevertheless, we observe visible changes to both the lamp body and the lamp window. On the lamp body inside the arms of the microwave cavity we observe brown marks in a repeating pattern indicating that standing waves in the cavity are creating "hot spots" within the lamp. On the inner surface of the $MgF_2$ window, we observe a thin film of deposited material opposite where the inner quartz tube approaches near the window. This film creates a faint reflection with variable color owing to the irregular thickness of the film (see Figure 5). X-ray spectroscopy scans of the window found no additional elements besides silicon and oxygen indicating that all significant contaminants have successfully been removed from the system and the film occurs due to deposition of quartz etched from the lamp body. Interestingly, Dong et al. found that $MgF_2$ did not significantly decrease its transmissivity over weeks of operating a 124 nm pulsed laser at intensities considerably higher than the UV lamp [7] implying that color centers likely will not contribute to long term degradation. This suggests that, in the absence of other contaminates, the long-term VUV output of the lamp may be dominated by a competition between etching and depositing of quartz onto the $MgF_2$ window surface during lamp operation. The rapid degradation of VUV lamp output observed in previous studies [8, 37] is therefore likely due to deposition of other contaminates or borosilicate lamp body material on the $MgF_2$ surface.

The lamp lifetime data was primarily taken for Xe*, where 147 nm output from the lamp is of concern. Because Kr* requires the output of 124 nm, where transmission efficiency of the $MgF_2$ is lower to begin with and may be more severely impacted by these etching and deposition effects, we cannot equivalently claim the same lifetime for the Kr* source yet. Nevertheless, after 166 hours of operation, the Kr* signal was found to be ~2/3 of the signal recorded in a separate, new copy of the lamp under similar conditions.

**Comparison to a radiofrequency-driven plasma discharge source**

To better contextualize lamp-based optical metastable excitation, we compare it to a radiofrequency (RF)-driven plasma discharge source. Figure 6 shows the Xe* and Kr* production for both metastable excitation methods as a function of input beam flux. Note that an external B-field was not applied for either optical or RF-driven metastable production recorded in Figure 6 although it is important to note that in addition to the B-field-enhanced metastable production reported here, B-field enhanced Kr* production using an RF-driven discharge source was also recently reported [25].

We find that the lamp excitation is equal to or more efficient than the RF source for all input fluxes. For the optical source, we observe qualitative behavior consistent with previous lamp-based optical excitation of a krypton beam [37]. At low pressures, the metastable flux increases approximately linearly with total beam flux. However, at higher input fluxes and thus higher background gas pressure, the metastable flux levels off and even decreases slightly. The decrease is likely a consequence of the mean free path for metastable atoms becoming comparable to the distance they travel before detection. Looking at Kr* production, the RF-driven production clearly demonstrates an optimal atom flux between $2\times10^{16}$ and $10^{17}$ atoms/s. Below and above this range, optical Kr* excitation is better by a factor of a few, indicating that optical Kr* excitation could prove extremely useful for measuring either small samples or achieving the highest flux for rapid sample analysis. The peak Kr* flux achieved with optical excitation is over a factor of 2 higher than with RF-driven excitation. For Xe* excitation, optical excitation is even more favorable at large input fluxes. Not only is the Xe* excitation efficiency overall typically an order of magnitude higher than for Kr* at the same input atom flux, but lamp-based optical excitation greatly exceeds RF-driven Xe* excitation efficiency for large input flux. The peak Xe* flux achieved with optical excitation is about 6 times greater than that achieved with RF-driven excitation and over 20 times greater than the Kr* flux achieved with RF-driven excitation. Combining the direct comparison with further enhancement of Xe* flux from an externally applied B-field demonstrates that xenon ATTA using lamp-based optical metastable excitation is a promising platform to explore improving upon current radioxenon detection techniques employed in the IMS.

The global trend of decreasing efficiency with increasing flux can be partially attributed to background gas collisions. This effect is not corrected for in Figure 6 and it affects the RF-driven plasma data more as the distance between excitation and detection regions is longer. An estimate of this bias can be computed for the krypton signals using the calculated Kr*-Kr elastic and metastable exchange cross section [38] by assuming that every collision results in the removal of Kr* from the probe volume. Performing this correction, we find that the corrected signal at the maximum flux would be enhanced by a factor of 2 and 3.4 for the lamp and RF signals respectively in the absence of background gas collisions. At smaller fluxes, the correction is negligible. For Xe, we estimate the same correction by assuming the ratio of Xe*-Xe to Xe-Xe cross-sections are equal to the equivalent ratio of Kr cross-sections since Xe*-Xe cross sections have not been previously measured or calculated. After correcting the observed data to remove this effect we estimate a lamp (RF) metastable excitation efficiency of $3.3\times10^{-4}$ ($6\times10^{-5}$) for Xe at a flux of $4\times10^{17}$ atoms per second, i.e., a factor of 1.6 (2.3) improvement in efficiency is expected in the absence of collisions. We expect that metastable production in a full ATTA system would be able to reduce the effect of background gas collisions due to the increased vacuum pumping speed and improved differential pumping compared to the vacuum system used for this work.

## Conclusions

The VUV optical source characterized in this work enables the generation of high-flux metastable krypton and xenon beams for ATTA with several advantages compared to state-of-the-art RF-driven sources. In particular, this lamp-based source alleviates problems due to cross-sample contamination for Kr and eliminates the minimum sample volume necessary in ATTA systems using traditional RF-discharge sources [8], all without any evidence of significant degradation on the 100-hour timescale.

Furthermore, our direct comparison between lamp-based optical excitation and RF-driven discharge excitation under nearly identical test conditions shows that our lamp-based source outperforms a traditional RF-driven discharge for both Kr* and Xe* production across the entire input flux range explored in this work. Comparing Xe* and Kr* production efficiency for both excitation methods, we show that Xe* production is approximately an order of magnitude more efficient than Kr*, except at the largest input fluxes tested where RF-discharge-based Xe* excitation is only slightly more efficient than for Kr*. Meanwhile, lamp-based optical Xe* excitation greatly exceeds RF-discharge excitation in this regime, making ATTA a promising technique for environmental detection of radioactive xenon isotopes. We also see potential room for improvement using thinner $MgF_2$ windows, particularly for the shorter krypton wavelength. Unlike Kr-ATTA, a practical limitation for Xe-ATTA measurements is likely total metastable flux rather than efficiency due to the relative abundance of atmospheric sample compared to the small amounts of Kr extracted from underground aquifers or polar ice. Since our maximum flux is likely limited by background gas collisions in this work, we believe that Xe* flux could be increased by a factor of 2 or more with the larger pumping speeds employed in typical ATTA instruments. As pointed out in other works [8, 10], adding more lamps to the system can also be used to increase metastable production in the beam, though we note that preliminary tests with two lamps have demonstrated non-trivial crosstalk that prevents signals from adding linearly, as naively expected.

Finally, the novel introduction of an external magnetic field to enhance lamp VUV output intensity and stability is both another source of potential improvement through optimization, and a broader technological advancement. External magnetic fields could be implemented in other VUV discharge lamps that are employed across a variety of other applications to achieve similar enhancements.

This work is supported by the U.S. Defense Threat Reduction Agency (DTRA) under Interagency Agreement DTRA1308144525 and by the U.S. DOE, Office of Science under contract DE-AC02-06CH11357. D. Koch acknowledges support from the U.S. Department of Energy, Office of Science, Office of Workforce Development for Teachers and Scientists (WDTS) under the Science Undergraduate Laboratory Internships Program (SULI).

# Figures

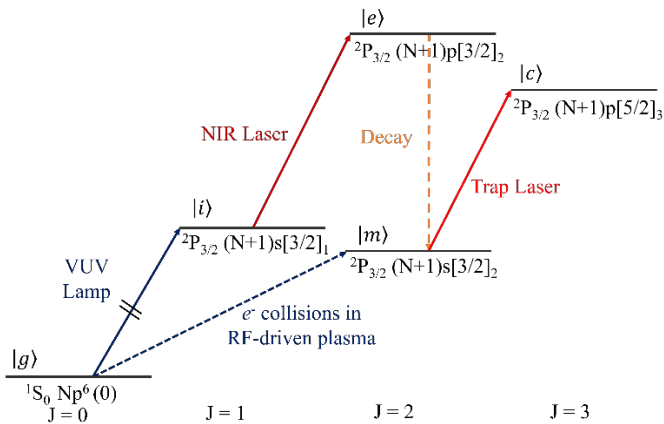

**Figure 1:** The shared iso-electronic energy level diagram relevant to krypton (N=4) and xenon (N=5) metastable production (not to scale). Arrows indicate the resonant pathways of the optical method, the non-resonant electron collision of the RF discharge source, and the optical cycling transition used for detection. State labels (|g>, |i>, |e>, |m>, |c>) are used to simplify description of the all-optical pumping scheme described in the text.

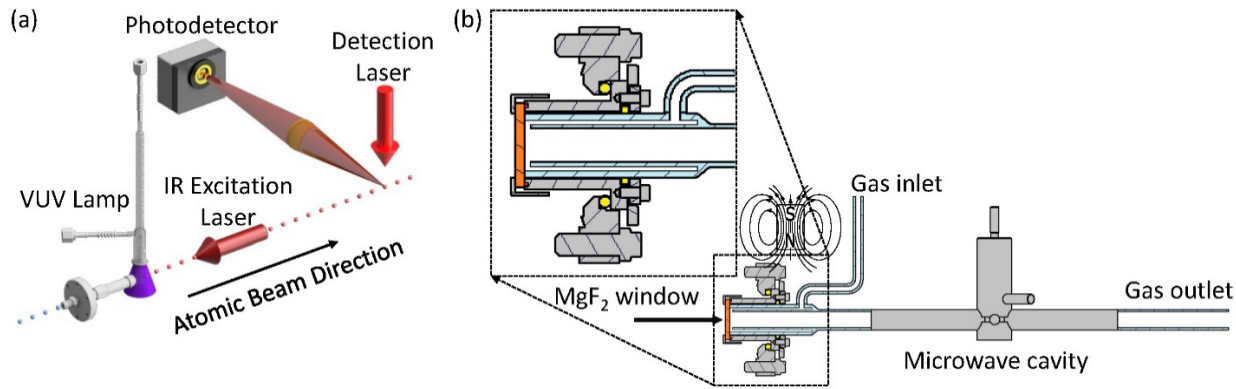

Figure 2: a) Experimental setup for optical excitation and detection of the atomic beam. b) Diagram of the flowing lamp and lamp-mounting fixture. (inset) A detailed view of the lamp-mounting fixture. A threaded cap presses the MgF$_2$ window (orange) against an indium gasket (red) to separate the vacuum system from the lamp gas. Steel screws (not shown in this view) attach the mounting fixture to the vacuum chamber and create a seal by compressing a Viton O-ring (yellow). The quartz lamp (light blue) is inserted and sealed by a Viton O-ring that is compressed with a steel plate screwed into the fixture. Plasma discharge is generated and tuned using a McCarroll-style microwave cavity [35] at 2.45 GHz. A neodymium rare earth magnet was introduced near the lamp window outside the vacuum chamber to enhance lamp output.

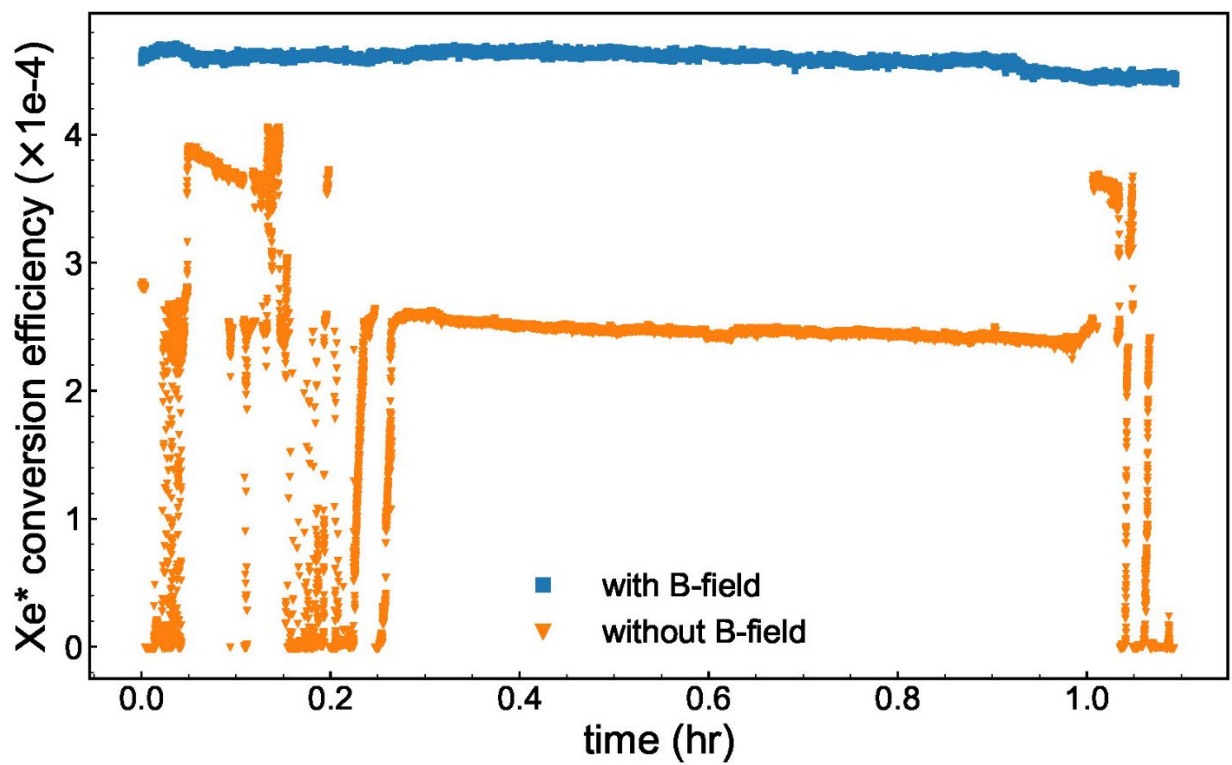

Figure 3: Observed metastable flux while operating the VUV lamp with (blue squares) and without (orange triangles) an applied magnetic field. Without the applied field, operating the lamp at high microwave power to achieve peak signal results in unstable operation and requires frequent adjustment to maintain optimum metastable excitation due to thermal effects. A reduction in microwave power allows the lamp to operate with stable metastable production but with decreased efficiency. Application of a magnetic field enables stable lamp operation with high efficiency.

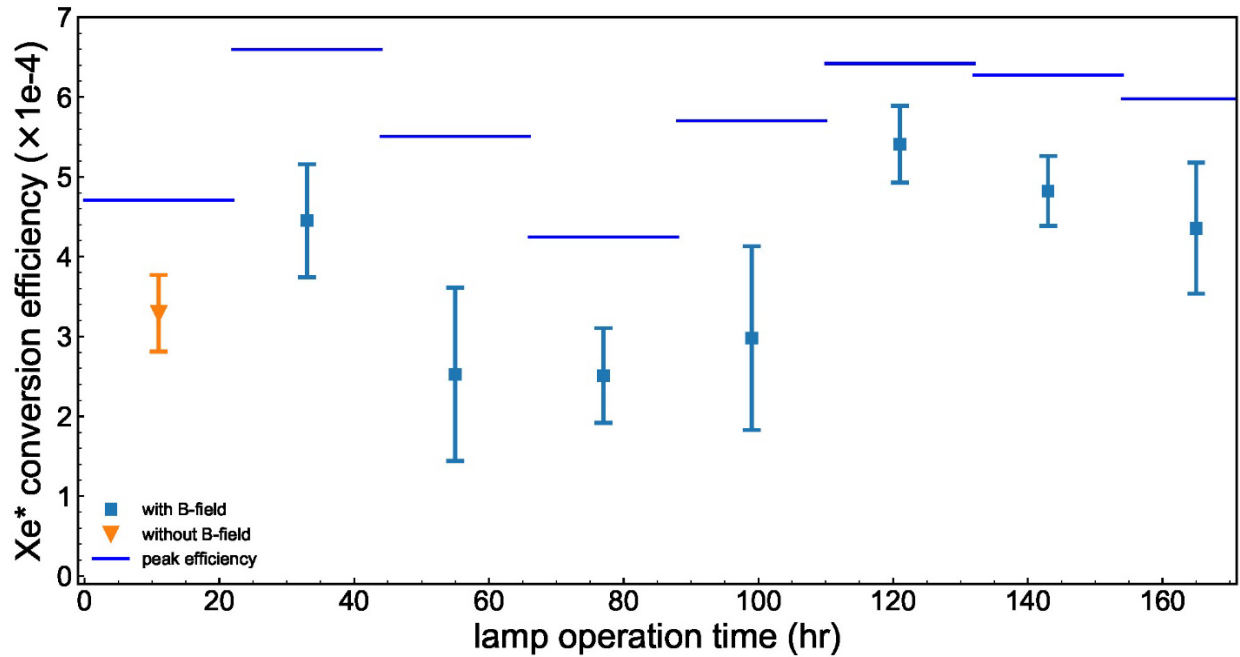

Figure 4: An evaluation of VUV lamp lifetime. We plot mean measured metastable conversion efficiency of xenon (points) versus lamp operation time. Error bars represent the standard deviation of observed values during stable lamp oeration (excluding plasma mode hops and periods of operator intervention). Systematic uncertainty in calculated efficiency (+/-10%) is limited by our determination of atomic beam temperatrue. Horizontal lines represent the peak observed conversion efficiency for the duration period within the width of the line.

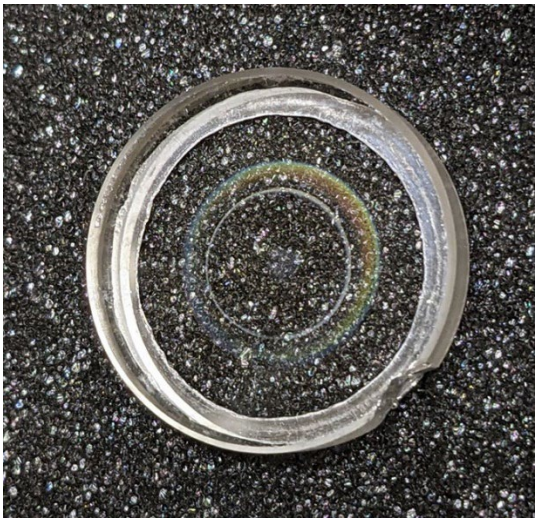

Figure 5: An image of a $MgF_2$ window that was removed from the VUV lamp after >100 hours of operation. The visible ring of deposited material was found to be silicon and oxygen from the quartz lamp body. The silvery ring of the leftover indium seal can be seen on the outer edge of the same face.

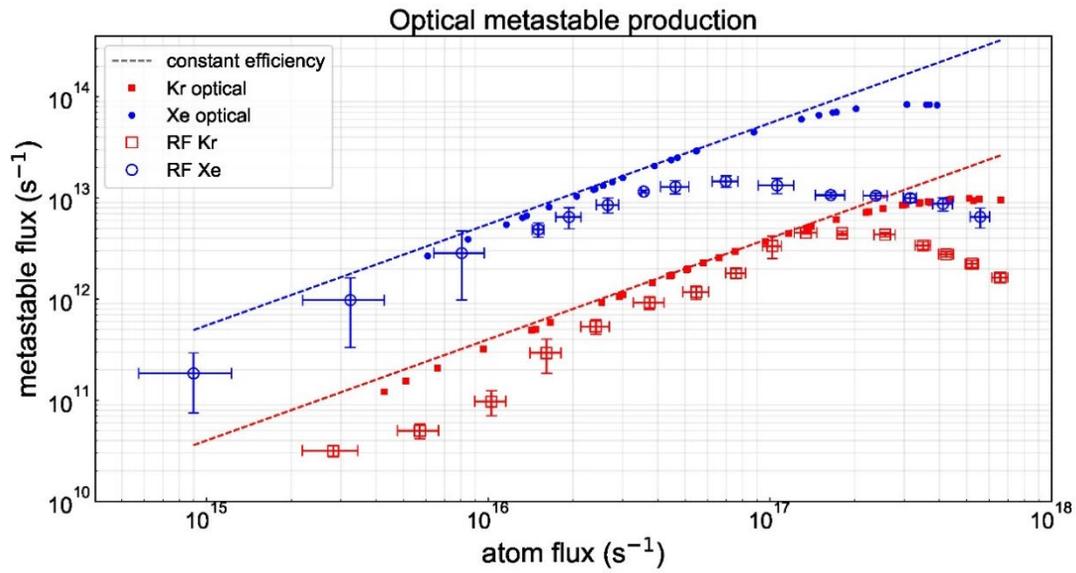

Figure 6: A comparison of lamp-based all-optical and RF-driven plasma metastable production for xenon and krypton atomic beams. Dashed lines represent constant metastable conversion efficiency for xenon (5.5e-4) and krypton (4e-5). External magnetic fields were not applied for this data.